\magnification=1200
\baselineskip=18truept
\def\bold1{{\bf 1}}

\def\ovlap{1}
\def\witt{2}
\def\geom{3}
\def\berry{4}
\def\odddim{5}
\def\jackiwlect{6}
\def\trunc{7}
\def\twod{8}

\line{\hfill RU-98-08}

\vskip 2truecm
\centerline{\bf Explicitly real form of the Wilson-Dirac matrix for
$SU(2)$.}

\vskip 1truecm
\centerline{Herbert Neuberger}
\vskip .5truecm
\centerline{\tt neuberg@physics.rutgers.edu}
\vskip 1truecm
\centerline {Department of Physics and Astronomy}
\centerline {Rutgers University, Piscataway, NJ 08855-0849}

\vskip 2truecm

\centerline{\bf Abstract}
\vskip 0.75truecm
The Wilson-Dirac matrix for $SU(2)$
with $I={1\over 2}$ fermions is written
in an explicitly real form. The basis change
relating it to the conventional form in
which the matrix has complex entries is
also given. Some applications are presented.

\vfill\eject
The Wilson-Dirac operator for $SU(2)$ with
fermions in the fundamental representation 
on a lattice of size $L^4$
is usually represented by an $8L^4 \times 8L^4$ complex
matrix. Below, I first present the
Wilson-Dirac operator as a real matrix of the same size.
Next, I write down the basis change relating it to the
complex form. I end with some comments. 

To construct the Wilson-Dirac operator we need, at each
lattice site, real linear combinations of elements
of an algebra generated by ${\bf 1},
\{\gamma_\mu ,~\mu=1,2,3,4\},~\{i\sigma_j,~j=1,2,3\}$,
acting on an $8\times 8$ dimensional space.
${\bf 1}$ is the $8\times 8$ unit matrix. 
The $\gamma_\mu$ and $\sigma_j$ (no $i$-factor) are
hermitian. 
The following relations define the needed 
seven matrices up to 
a choice of basis:
$$
[\gamma_\mu , i\sigma_j ]=0 ~,~~~~i\sigma_j i\sigma_k =
-\epsilon_{jkl} i\sigma_l ~,~~~~(i\sigma_j)^2=-\bold1~,~~~~
\{\gamma_\mu ,\gamma_\nu\}=2\delta_{\mu\nu}.\eqno{(1)}$$

The real form of the
Wilson-Dirac operator is now defined
by exhibiting seven real $8\times 8$ 
matrices obeying (1). To distinguish these matrices
from the more standard, complex, versions I denote
them by the corresponding capital
letters. For practical purposes I looked for a form
in which the $\Gamma_\mu$ matrices are sparse, have
simple numerical entries, and 
have an off-diagonal structure characteristic of a
chiral basis. 

The $\Gamma_\mu$ have the following structure in terms of 
$4\times 4$ blocks:
$$
\Gamma_\mu =\pmatrix{0&\rho_\mu\cr\rho_\mu^T & 0},
\eqno{(2)}$$
where $\rho_\mu^T$ denotes the transpose of $\rho_\mu$. 
The $i\Sigma_j$ have the following structure in
$4\times 4$ blocks:
$$
i\Sigma_j =\pmatrix{s_j & 0\cr 0 & s_j}~,~~~~s_j=-s_j^T~,~~~~
s_j^2=-\bold1.\eqno{(3)}$$

The explicit $4\times 4$ matrices we need are listed
below:
$$
\rho_1 =\pmatrix{0&0&-1&0\cr 
                 0&0&0&-1\cr
                 1&0&0&0\cr
                 0&1&0&0}~
\rho_2 =\pmatrix{0&0&0&-1\cr 
                 0&0&1&0\cr
                 0&-1&0&0\cr
                 1&0&0&0}~
\rho_3 =\pmatrix{0&1&0&0\cr 
                 -1&0&0&0\cr
                 0&0&0&-1\cr
                 0&0&1&0}$$
$$
\rho_4 =\pmatrix{-1&0&0&0\cr 
                 0&-1&0&0\cr
                 0&0&-1&0\cr
                 0&0&0&-1}$$
$$
s_1 =\pmatrix{0&1&0&0\cr 
                 -1&0&0&0\cr
                 0&0&0&1\cr
                 0&0&-1&0}~
s_2 =\pmatrix{0&0&0&1\cr 
                 0&0&1&0\cr
                 0&-1&0&0\cr
                 -1&0&0&0}~
s_3 =\pmatrix{0&0&-1&0\cr 
                 0&0&0&1\cr
                 1&0&0&0\cr
                 0&-1&0&0}\eqno{(4)}$$

The above representation is related by a
basis change to the usual, complex, version.
More precisely, the $\gamma_\mu$ matrices are
taken in a chiral basis as complex $4\times 4$
matrices:
$$ 
\gamma_k = \sigma_1\otimes\sigma_k ,~~~~k=1,2,3,~~~~~~~~~~~~
\gamma_4 = \sigma_2 \otimes\bold1 ,
\eqno{(5)}$$
where the 
$2\times 2$
matrices $\sigma_k$ are the standard
Pauli matrices acting on two dimensional
complex space. The Pauli matrices are also 
chosen for the action on the group index.
I replace the single four valued spinorial
index in the usual representation by two
two-valued indices matching the direct products
of equation (5). Thus, a fermion field
in the usual representation carries four indices:
The first three indices take two values 
and label the eight dimensional space per site.
The last index takes $L^4$ values and labels the
site. The first pair of indices are Dirac and the
third index is $SU(2)$-color. 

Acting on these fermion fields we define the
unitary matrix $\Phi$:
$$
\Phi = {1\over 2} \alpha\otimes\left [ \alpha\otimes\beta +
\beta \otimes \alpha +\beta\otimes
\beta - \alpha\otimes\alpha\right ]\otimes\bold1 ,\eqno{(6)}$$
where $\alpha$ and $\beta$ are given by
$$
\alpha=\pmatrix{1&0\cr 0& i\cr}~,~~~~\beta=\pmatrix{0&1\cr i& 0\cr}.
\eqno{(7)}
$$

The four $\gamma_\mu$ matrices 
turn into the $\Gamma_\mu$ matrices under conjugation.
The latter are written below 
using three two valued indices rather than
the one eight valued index employed equations (3,4): 
$$
\eqalign{
\Phi ( \gamma_1 \otimes\bold1\otimes\bold1 )
\Phi^\dagger = 
\Gamma_1 \otimes\bold1 &\equiv
~-\varepsilon\otimes ~\varepsilon\otimes\bold1
\otimes\bold1 \cr
\Phi ( \gamma_2 \otimes\bold1\otimes\bold1 )\Phi^\dagger =
\Gamma_2 \otimes\bold1&\equiv ~
-\varepsilon\otimes\sigma_1\otimes\varepsilon
\otimes\bold1\cr 
\Phi ( \gamma_3 \otimes\bold1\otimes\bold1 ) \Phi^\dagger =
\Gamma_3 \otimes\bold1&\equiv 
~~~\varepsilon\otimes\sigma_3\otimes\varepsilon
\otimes\bold1\cr
\Phi ( \gamma_4 \otimes\bold1\otimes\bold1 ) \Phi^\dagger =
\Gamma_4 \otimes\bold1&\equiv 
-\sigma_1\otimes\bold1\otimes\bold1
\otimes\bold1,\cr}
\eqno{(8)}$$
where
$$\varepsilon =-i\sigma_2 =\pmatrix{0&-1\cr 1& 0},~~~~
\varepsilon^2 =-\bold1.\eqno{(9)}$$

Similarly, the $i\sigma_j$ matrices become under conjugation the
$i\Sigma_j$ matrices of equations (3,4):
$$
\eqalign{
\Phi ( \bold1\otimes\bold1\otimes(i\sigma_1)\otimes\bold1 )
\Phi^\dagger = 
i\Sigma_1 \otimes\bold1 &\equiv
-\bold1\otimes \bold1\otimes ~\varepsilon
\otimes\bold1 \cr
\Phi ( \bold1\otimes\bold1\otimes(i\sigma_2)\otimes\bold1 )
\Phi^\dagger = 
i\Sigma_2 \otimes\bold1 &\equiv
-\bold1\otimes \varepsilon \otimes\sigma_1 
\otimes\bold1 \cr
\Phi ( \bold1\otimes\bold1\otimes(i\sigma_3)\otimes\bold1 )
\Phi^\dagger = 
i\Sigma_3 \otimes\bold1 &\equiv ~~
\bold1\otimes \varepsilon \otimes\sigma_3 
\otimes\bold1 .\cr}\eqno{(10)}$$

The reality of the Wilson-Dirac operator
should be useful in simulations of $SU(2)$
gauge theories. From eq. (4) we see that the action
of the $\rho_\mu$ matrices amounts to permutations plus sign
switches. The projector combinations one encounters
in the Wilson-Dirac operator, $1\pm \Gamma_\mu$, are
easily efficiently implemented because the matrices
$\rho_{1,2,3}$ are antisymmetric and $\rho_4$ is symmetric. 
When compared to QCD one would expect the computational
gain in fermion manipulations (for example in the inversion
of the Wilson-Dirac matrix) when restricting the gauge group to $SU(2)$,
but staying with a complex representation,  
to be by a factor of about ${3\over 2}$. 
I have implemented the above real 
form in a parallelized high level Fortran 90 code
and a comparison has given a gain
of about 6.
A beta version of the
package I used is available upon request. 
A real form of the Wilson-Dirac operator can be
obtained for any non-integral isospin, but not
for integral isospin. 

The above basis is particularly useful for
the overlap [\ovlap]. 
The matrix $\gamma_5$ is 
diagonal before and after
the $\Phi$ transformation. 
The Hamiltonian matrix used in the
overlap, $H$, is given by $\Gamma_5 D_W$,
where $D_W$ is the Wilson-Dirac
matrix.  $H$ is both real and hermitian, 
hence
symmetric. All the eigenvectors of 
$H$ can be chosen
real. Thus, the overlap itself is real. This means that a theory
representing a single Weyl doublet in interaction with the $SU(2)$
gauge field, has a real fermion determinant on the 
lattice just like in the continuum. Under gauge transformations sign
switches can occur. This is how Witten's anomaly [\witt]
will be reflected 
on the lattice.
On the basis of [\geom] 
these sign changes
can be associated with degeneracies 
of  $H$ 
of codimension 2 in this case (conical degeneracies [\berry]). 
Such degeneracies are known to occur in the continuum by Witten's
argument [\witt]. 

What we need here is made explicit in what follows:
Consider first the massless Dirac operator, in Euclidean continuous
space, made hermitian and real. We take space as a torus of size $l$
in each direction. Let us label the 
space-time points by four vectors $\xi$.
Fixing $\xi_4$, we have a three torus which can be 
mapped into $SU(2)$ with unit winding [\odddim]:
$$
G (\xi_1 ,\xi_2 ,\xi_3) = 
{{-1 +\sum_{\mu=1}^3 [1-\cos ({{2\pi}\over l} \xi_\mu )]
+i\sum_{\mu=1}^3 \sigma_\mu \sin ({{2\pi}\over l} \xi_\mu )}\over
{\sqrt { \{ -1 + \sum_{\mu=1}^3 [1-\cos ({{2\pi}\over l} \xi_\mu )]
\}^2 +
\sum_{\mu=1}^3 [ \sin ({{2\pi}\over l} \xi_\mu )]^2 }}}. \eqno{(11)}$$
This choice preserves the symmetry under discrete rotations compatible
with the three torus. 
Now, following a formula due to Goldstone, presented in [\jackiwlect],
we introduce a gauge transformation depending on all four coordinates:
$$
{\cal G} (\xi ) = e^{i{\pi\over l} \xi_4 \sigma_3 }
G^\dagger (\xi_1 ,\xi_2 ,\xi_3 )  e^{-i{\pi\over l} \xi_4 \sigma_3 }
G (\xi_1 ,\xi_2 ,\xi_3 ).\eqno{(12)}$$
This map cannot be deformed to unity. 
[\witt] shows 
that as the vector potential associated with ${\cal G}$
is deformed to zero a conical singularity will occur in the 
massless Dirac operator somewhere along the way. 

It is trivial to put ${\cal G}$ on the lattice, and to again
define a link by link interpolation to unit link variables
everywhere. Now, in addition to the deformation parameter,
we also have the mass 
parameter $m$ in $D_W$ (alternatively known
as the hopping parameter $\kappa$) 
at our disposal. To be definite, we 
choose $m=-1$ (corresponding to
$\kappa={1\over 6}$) 
and start increasing it towards zero. For positive $m$ (corresponding to $\kappa < {1\over
8}$), 
$H$ has a gap at zero [\ovlap] 
and nothing can happen. However,
the smoothness of the gauge 
configuration in (13) implies that in the limit
of infinite $L$ the continuum conical singularity must appear at zero $m$.
It cannot come from nowhere, so it must be somewhere in the
plane spanned by the real deformation parameter and negative $m$.

For $SU(2)$, 
the reality of the Wilson-Dirac operator also extends
to truncations of the overlap, which are almost
massless lattice versions of continuum
vector-like gauge theories [\trunc].
In the context of chiral gauge theories we recall that 
while $SU(2)$ gauge theory with a single Weyl fermion of isospin
${1\over 2}$ is inconsistent because of the global anomaly, 
if the isospin is ${3\over 2}$, the problem disappears [\witt].
Taking one Weyl fermion 
of isospin ${3\over 2}$ is probably the
simplest consistent four dimensional
chiral model 
with a real fermion
determinant. 
The reality is similar to the 11112 
model in two dimensions [\twod] which
yielded a successful test of the overlap.
On the other hand, the four dimensional
model has not been exactly solved in the
continuum and its properties are unknown.

{\bf Acknowledgment:} This research was 
supported in part by the DOE under grant \#
DE-FG05-96ER40559. Thanks are due to Y. Kikukawa and R. Narayanan 
for discussions at earlier stages of this research. 

\vskip 2truecm

{\bf  References:}

\vskip .5truecm

\item{[\ovlap]} R. Narayanan, H. Neuberger, Phys. Lett. B302 (1993) 62;
Nucl. Phys. B443 (1995) 305; Phys. Rev. Lett. 71 (1993) 3251.
\item{[\witt]} E. Witten, Phys. Lett. 117B (1982) 324.
\item{[\geom]} H. Neuberger, hep-lat/9802033.
\item{[\berry]} M. V. Berry, Proc. R. Lond. A392 (1984) 45.
\item{[\odddim]} Y. Kikuakwa, H. Neuberger, 
Nucl. Phys. B513 (1998) 735.
\item{[\jackiwlect]} R. Jackiw, in ``Relativity, 
groups and topology II'', Les Houches,
1983, eds. B. S. Dewitt and R. Stora (North-Holland, Amsterdam, 1984).
\item{[\trunc]} H. Neuberger, Phys. Rev. D57 (1998) 5417; 
Y. Kikukawa, H. Neuberger, A. Yamada,
hep-lat/9712022, Nucl. Phys. B, to appear.
\item{[\twod]} 
Y. Kikukawa, R. Narayanan, H. Neuberger, 
Phys. Lett. B508 (1997) 105; 
Phys. Rev. D57 (1998) 1233; 
H. Neuberger, R. Narayanan, Phys. Lett. B402 (1997) 320.

\vfill\eject\end